\documentstyle[aps]{revtex}

\newcommand{\balpha}{{\mbox{\boldmath$\alpha$}}}
\newcommand{\bnabla}{{\mbox{\boldmath$\nabla$}}}
\newcommand{\bxi}{{\mbox{\boldmath$\xi$}}}
\newcommand{\be}{\begin{eqnarray}}
\newcommand{\ee}{\end{eqnarray}}
\newcommand{\la}{\langle}
\newcommand{\ra}{\rangle}

\newcommand{\bfx}{{\bf x}}
\newcommand{\bfy}{{\bf y}}
\newcommand{\bfr}{{\bf r}}
\newcommand{\bfa}{{\bf a}}
\newcommand{\bfR}{{\bf R}}
\newcommand{\bfP}{{\bf P}}
\newcommand{\bfK}{{\bf K}}
\newcommand{\bfA}{{\bf A}}
\newcommand{\bfD}{{\bf D}}
\newcommand{\bfX}{{\bf X}}
\newcommand{\bfY}{{\bf Y}}
\newcommand{\bfZ}{{\bf Z}}

\newcommand{\bfz}{{\bf z}}
\newcommand{\bfn}{{\bf n}}
\newcommand{\bfk}{{\bf k}}
\newcommand{\bfp}{{\bf p}}
\newcommand{\veps}{\varepsilon}

\newcommand{\eps}{\epsilon}

\begin{document}
\title
{ QED theory of the nuclear recoil effect on the atomic $g$ factor }
\author{V. M. Shabaev$^{1,2}$}

\address
{$^1$Department of Physics, St.Petersburg State University,
Oulianovskaya 1, Petrodvorets, St.Petersburg 198504, Russia\\
$^2$Gesellschaft f\"ur Schwerionenforschung, 64291 Darmstadt,
Germany}
\maketitle
\begin{abstract}
The quantum electrodynamic theory of the nuclear recoil
effect on the atomic $g$ factor  to all orders in $\alpha Z$ 
and to first order in $m/M$ is formulated. 
The complete $\alpha Z$-dependence formula
for the recoil correction to the bound-electron $g$ factor
in a hydrogenlike atom is derived. This formula is used
to calculate the recoil correction to the bound-electron
$g$ factor in the order $(\alpha Z)^2 m/M$ for an arbitrary
state of a hydrogenlike atom.
\end{abstract}

\begin{center}
{\bf I.$\;$INTRODUCTION}
\end{center}

High precision direct measurements of the bound-electron
$g$ factor in low-$Z$ hydrogenlike ions, which are performed by 
a GSI - Universit\"at Mainz collaboration
\cite{gsi1,gsi2},
have triggered a great interest to the calculations
of QED and nuclear corrections to this effect
\cite{blu97,per97,bei00c,bei00a,bei00b,kar00a,kar00b,kar01,cza01,fau00}.
It is expected that in nearest future the collaboration
will extend these measurements to high-$Z$  systems.
The accurate measurements of the $g$ factor in heavy ions
would provide a good possibility for a test of the 
magnetic sector of QED in a strong Coulomb field.
To date, investigations of the bound-electron
$g$ factor in heavy ions were performed only by an
 indirect method \cite{sha98a}
which is based on using the fact that
the bound-electron $g$ factor can be expressed in terms
of the transition probability between the hyperfine
splitting components \cite{see98,win99}.
In particular, in the case of hydrogenlike bismuth,
it was found that including the QED correction to the
$g$ factor is necessary to obtain agreement between theory
and experiment (see \cite{win99,sha00} for details).
However, the indirect method can not provide so high accuracy
 as the direct method if the latter is extended to high-$Z$
systems.

It is known (see, e.g., \cite{sha00,moh98}) that calculations
of heavy ions have to be performed  without any expansion
in the papameter $\alpha Z$. In particular, it means that,
to zeroth-order approximation, the electron in a hydrogenlike
atom must be described by the Dirac equation with the Coulomb
potential induced by the nucleus.
For the point-charge nucleus, the bound-electron $g$ factor
in a hydrogenlike atom is easily calculated analytically
(see, e.g., \cite{zap79}).
 The one-loop QED corrections to
the $1s$ $g$ factor were calculated to all orders in
$\alpha Z$ in \cite{blu97,per97,bei00c,bei00a}
(see also \cite{bei00b}). The finite nuclear size correction
can be found by solving the Dirac equation for an
extended nucleus \cite{per97,sha98a}. For low- and middle-$Z$
atoms, this correction can also be found by non-relativistic
and approximate relativistic formulas \cite{kar00a,gla01}.
The nuclear recoil correction to the $1s$ $g$ factor
was evaluated only to the lowest order in $\alpha Z$
\cite{fau70a,fau70b,gro70,gro71} (see also \cite{clo71,fau00}).
The methods developed in \cite{fau70a,fau70b,gro70,gro71}
 are suitable for 
calculations to a few lowest orders in the parameter
$\alpha Z$ and are not intended for calculations of the
recoil effect to all orders in $\alpha Z$.
Therefore, first of all we need to develop a new method
which would provide a systematic treatment of
 the recoil effect on the $g$ factor to all orders 
in $\alpha Z$ and to first order in $m/M$ ($m$ is the
electron mass and $M$ is the nucleus mass).
Such a method is proposed in the present paper.
This method is valid for the general case of
 a many-electron atom.
Apllieng the method to a hydrogenlike atom,
the complete $\alpha Z$-dependence formula
for the recoil correction to the $g$ factor is derived.
To the lowest relativistic order ($\sim (\alpha Z)^2 m/M$), 
this formula is
used to calculate analytically
the recoil correction to the $g$ factor in the case
of  an arbitrary state of a hydrogenlike atom.
For the $1s$ state, the result obtained in this paper
coincides with the related result obtained previously
in \cite{fau70a,fau70b,gro70,gro71}.

The relativistic units ($\hbar=c=1$) and the Heaviside
charge unit ($\alpha =e^2/(4\pi),\;\;e<0$)
are used in the paper.

\begin{center}
{\bf II.$\;$QUANTUM ELECTRODYNAMICS TO FIRST ORDER IN 
$m/M$ BEYOND THE FURRY PICTURE}
\end{center}

To derive the Hamiltonian of the atom to first order
in $m/M$ and to lowest order in $\alpha$ (but to all
orders in $\alpha Z$), we 
start with the relativistic quantum mechanical treatment
of the electrons and the non-relativistic treatment of the nucleus.
For simplicity, we consider the nucleus as a spinless
particle. The electrons and the nucleus interact with each other,
with the quantized electromagnetic field, and with the
 classical homogeneous magnetic field, ${\bf A}_{\rm cl}(\bfr)
=[{\bf {\cal H}}\times {\bfr}]/2$. 
In the Coulomb gauge and the Schr\"odinger 
representation, the Hamiltonian of the system 
to first order in the interaction with $\bfA_{\rm cl}(\bfr)$ is
\begin{eqnarray} \label{eq1}
H&=&\sum_{i=1}^{N'}[\balpha_i\cdot \bfp^{(e)}_{i}
+\beta_i m+V(\bfr^{(e)}_i-\bfr^{(n)})
-e\balpha_i\cdot {\bf A}(\bfr^{(e)}_i)]\nonumber\\
&&+\frac{1}{2}\sum_{i\ne k}\frac{\alpha}{|\bfr^{(e)}_i
-\bfr^{(e)}_k|}+\frac{1}{2M}[\bfp^{(n)}-|e|Z{\bf A}(\bfr^{(n)})]^2
\nonumber\\
&&+\frac{1}{2}
\int d{\bf x}\;
[{\bf {\cal E}}_{t}^{2}({\bf x})+{\bf {\cal H}}^{2}({\bf x})]
\nonumber\\
&&-e\sum_{i=1}^{N'}\balpha_i\cdot {\bf A}_{\rm cl}(\bfr^{(e)}_i)
-\frac{|e|Z}{M}\bfp^{(n)}\cdot {\bf A}_{\rm cl}(\bfr^{(n)})\,,
\end{eqnarray}
where $N'$ is the total number of the positive and negative
energy state electrons, $\bfp^{(e)}_i$ and $\bfp^{(n)}$ are the
electron and nucleus momentum operators, respectively,
$V(\bfr^{(e)}_i-\bfr^{(n)})$ is the operator of the 
electron-nucleus Coulomb interaction, ${\bf A}(\bfr)$
is the vector potential of the quantized electromagnetic field,
${\bf {\cal H}}=\bnabla\times{\bf A}$, and
${\bf {\cal E}}_t=-{\partial}{\bf A}/{\partial}t$. 
The last two terms in equation (\ref{eq1}) describe
the interaction of the electrons and the nucleus
with the homogeneous magnetic field.
We introduce the center-of-mass variables for the
electron-nucleus subsystem
\be
{\bf R}&=&\frac{1}{M+N'm}\Bigl(
M\bfr^{(n)}+m\sum_{k=1}^{N'}\bfr^{(e)}_k\Bigr)\,,\\
\bfr_i&=&\bfr_i^{(e)}-\bfr^{(n)}\,.
\ee
From these equations we derive
\be \label{eq4}
\bfr_i^{(e)}&=&\bfr_i+{\bf R}-\frac{m}{M+N'm}\sum_{k=1}^{N'}\bfr_k\,,\\
\bfr^{(n)}&=&{\bf R}-\frac{m}{M+N'm}\sum_{k=1}^{N'}\bfr_k\,.
\label{eq5}
\ee
For the corresponding momentum operators
(${\bfP}=-i\bnabla_{\bfR}$, $\bfp_i=-i\bnabla_{\bfr_i}$),
we obtain
\be \label{eq6}
\bfp_i^{(e)}&=&\bfp_i+\frac{m}{M+N'm}\bfP\,,\\
\bfp^{(n)}&=&\frac{M}{M+N'm}\bfP-\sum_{k=1}^{N'}\bfp_k\,.
\label{eq7}
\ee
To keep in the Hamiltonian (\ref{eq1}) the terms of zeroth
and first orders in $m/M$, we can replace equations
(\ref{eq4})-(\ref{eq7}) by the following ones
\be  \label{eq8}
\bfr_i^{(e)}&=&\bfr_i+{\bf R}-\frac{m}{M}\sum_{k=1}^{N'}\bfr_k\,,\\
\bfr^{(n)}&=&{\bf R}-\frac{m}{M}\sum_{k=1}^{N'}\bfr_k\,,
\label{eq9}\\
\bfp_i^{(e)}&=&\bfp_i+\frac{m}{M}\bfP\,, \label{eq10}\\
\bfp^{(n)}&=&\bfP-\sum_{k=1}^{N'}\bfp_k\,.
\label{eq11}
\ee
With these substitutions, the individual terms
in equation (\ref{eq1}) are transformed as
\be \label{eq12}
\sum_{i=1}^{N'}\balpha_i\cdot\bfp_i^{(e)}&=&
\sum_{i=1}^{N'}\balpha_i\cdot\bfp_i+\frac{m}{M}
\sum_{i=1}^{N'}\balpha_i\cdot\bfP\,,\\
\sum_{i=1}^{N'}V(\bfr_i^{(e)}-\bfr^{(n)})&=&
\sum_{i=1}^{N'}V(\bfr_i)\,, \label{eq13}\\
-e\sum_{i=1}^{N'}\balpha_i\cdot \bfA(\bfr_i^{(e)})&=&
-e\sum_{i=1}^{N'}\balpha_i\cdot \bfA\Bigl(\bfr_i+\bfR
-\frac{m}{M}\sum_{k=1}^{N'}\bfr_k\Bigr)\nonumber\\
&=&-e\sum_{i=1}^{N'}\balpha_i\cdot \bfA(\bfr_i+\bfR)
\nonumber\\
&&+e\frac{m}{M}\Bigl[\Bigl(\sum_{k=1}^{N'}\bfr_k\cdot
\frac{\partial}{\partial\bxi}\Bigr)\Bigl(\sum_{i=1}^{N'}
\balpha_i\cdot\bfA(\bxi)\Bigr)\Bigr]_{\bxi=\bfr_i+\bfR}\,, 
\label{eq14}\\
\frac{1}{2}\sum_{i\ne k}\frac{\alpha}{|\bfr^{(e)}_i-\bfr^{(e)}_k|}
&=&\frac{1}{2}\sum_{i\ne k}\frac{\alpha}{|\bfr_i-\bfr_k|}\,,
\label{eq15}\\
\frac{1}{2M}[\bfp^{(n)}-|e|Z\bfA(\bfr^{(n)})]^2&=&
\frac{\bfP^2}{2M}-\frac{1}{M}\bfP\cdot\sum_{k=1}^{N'}\bfp_k
+\frac{1}{2M}\Bigl(\sum_{k=1}^{N'}\bfp_k\Bigr)^2\nonumber\\
&&-\frac{|e|Z}{M}\bfP\cdot\bfA(\bfR)
+\frac{|e|Z}{M}\Bigl(\sum_{k=1}^{N'}\bfp_k\Bigr)\cdot\bfA(\bfR)
+\frac{e^2 Z^2}{2M}\bfA^2(\bfR)\,,
\label{eq16}\\
-e\sum_{i=1}^{N'}\balpha_i\cdot\bfA_{\rm cl}(\bfr^{(e)}_i)
&=&-e\sum_{i=1}^{N'}\balpha_i\cdot\bfA_{\rm cl}\Bigl(\bfr_i
+\bfR-\frac{m}{M}\sum_{k=1}^{N'}\bfr_k\Bigr)\nonumber\\
&=&-e\sum_{i=1}^{N'}\balpha_i\cdot\bfA_{\rm cl}(\bfr_i)
-e\sum_{i=1}^{N'}\balpha_i\cdot\bfA_{\rm cl}(\bfR)
\nonumber\\
&&+e\frac{m}{M}\sum_{i=1}^{N'}\balpha_i\cdot\sum_{k=1}^{N'}
\bfA_{\rm cl}(\bfr_k)\,,
\label{eq17}\\
-\frac{|e|Z}{M}\bfp^{(n)}\cdot \bfA_{\rm cl}(\bfr^{(n)})&=&
-\frac{|e|Z}{M}\bfP\cdot\bfA_{\rm cl}(\bfR)
+\frac{|e|Z}{M}\sum_{k=1}^{N'}\bfp_k\cdot\bfA_{\rm cl}(\bfR)\,.
\label{eq18}
\ee
Here we have disregarded the terms of second and higher
orders in $m/M$.
Because the second term in the right hand side of equation 
(\ref{eq14}) contains the factor $e\frac{m}{M}$,
to first order in $m/M$,  it can contribute
only to first and higher orders in $\alpha$. Therefore,
to the order under consideration (to zeroth order in $\alpha$,
to first order in $m/M$, and to all orders in $\alpha Z$)
this term can be omitted. We obtain
\be \label{eq19}
H&=&\sum_{i=1}^{N'}[\balpha_i\cdot \bfp_{i}
+\beta_i m+V(\bfr_i)]+\frac{1}{2}\sum_{i\ne k}\frac{\alpha}{|\bfr_i
-\bfr_k|} \nonumber\\
&& +\frac{1}{2}\int d{\bf x}\;
[{\bf {\cal E}}_{t}^{2}({\bf x})+{\bf {\cal H}}^{2}({\bf x})]
-e\sum_{i=1}^{N'}\balpha_i\cdot \bfA(\bfr_i+\bfR)
\nonumber\\
&&+\frac{\bfP^2}{2M}
+\frac{1}{2M}\Bigl(\sum_{k=1}^{N'}\bfp_k\Bigr)^2
+\frac{m}{M}\sum_{i=1}^{N'}\balpha_i\cdot\bfP\nonumber\\
&&-\frac{1}{M}\bfP\cdot\sum_{k=1}^{N'}\bfp_k
-\frac{|e|Z}{M}\bfP\cdot\bfA(\bfR)\nonumber\\
&&+\frac{|e|Z}{M}\Bigl(\sum_{k=1}^{N'}\bfp_k\Bigr)\cdot\bfA(\bfR)
+\frac{e^2 Z^2}{2M}\bfA^2(\bfR)\nonumber\\
&&-e\sum_{i=1}^{N'}\balpha_i\cdot\bfA_{\rm cl}(\bfr_i)
-e\sum_{i=1}^{N'}\balpha_i\cdot\bfA_{\rm cl}(\bfR)\nonumber\\
&&+e\frac{m}{M}\sum_{i=1}^{N'}\balpha_i\cdot\sum_{k=1}^{N'}
\bfA_{\rm cl}(\bfr_k)
-\frac{|e|Z}{M}\bfP\cdot\bfA_{\rm cl}(\bfR)
+\frac{|e|Z}{M}\sum_{k=1}^{N'}\bfp_k\cdot\bfA_{\rm cl}(\bfR)\,.
\ee
In the second quantized form, this Hamiltonian can be written as
\be \label{eq20}
H&=&H_0+\sum_{s=1}^{13}H_{\rm int}^{(s)}\,,\\
H_0&=&\int d\bfx\;\psi^{\dag}(\bfx)[-i\bnabla\cdot\balpha
+\beta m+V(\bfx)]\psi(\bfx)-\frac{1}{2M}\int d\bfx\;\phi^{\dag}
\Delta_{\bfx}\phi(\bfx)\nonumber\\
&&+\frac{1}{2}\int d{\bf x}\;
[{\bf {\cal E}}_{t}^{2}({\bf x})+{\bf {\cal H}}^{2}({\bf x})]\,,
\label{eq21}\\
H_{\rm int}^{(1)}&=&\frac{\alpha}{2} \int d\bfx\,d\bfy\;
\frac{\psi^{\dag}(\bfx)\psi(\bfx)\psi^{\dag}(\bfy)
\psi(\bfy)}{|\bfx-\bfy|}\,,
\label{eq22}\\
H_{\rm int}^{(2)}&=&-e\int d\bfx\,d\bfy\;
\psi^{\dag}(\bfx)\balpha\psi(\bfx)
\bfA(\bfx+\bfy)\phi^{\dag}(\bfy)\phi(\bfy)\,,
\label{eq23}\\
H_{\rm int}^{(3)}&=&\frac{1}{2M}\int d\bfx\;
\psi^{\dag}(\bfx)(-i\bnabla_{\bfx})\psi(\bfx)
\int d\bfy\;
\psi^{\dag}(\bfy)(-i\bnabla_{\bfy})\psi(\bfy)\,,
\label{eq24}\\
H_{\rm int}^{(4)}&=&\frac{|e|Z}{M}\int d\bfx\;
\psi^{\dag}(\bfx)(-i\bnabla_{\bfx})\psi(\bfx)
\int d\bfy\;\bfA(\bfy)
\phi^{\dag}(\bfy)\phi(\bfy)\,,
\label{eq25}\\
H_{\rm int}^{(5)}&=&\frac{e^2 Z^2}{2M}
\int d\bfy\;\bfA^2(\bfy)
\phi^{\dag}(\bfy)\phi(\bfy)\,,
\label{eq26}\\
H_{\rm int}^{(6)}&=&\frac{m}{M}\int d\bfx\;
\psi^{\dag}(\bfx)\balpha\psi(\bfx)
\int d\bfy\;
\phi^{\dag}(\bfy)(-i\bnabla_{\bfy})\phi(\bfy)\,,
\label{eq27}\\
H_{\rm int}^{(7)}&=&-\frac{1}{M}\int d\bfx\;
\psi^{\dag}(\bfx)(-i\bnabla_{\bfx})\psi(\bfx)
\int d\bfy\;
\phi^{\dag}(\bfy)(-i\bnabla_{\bfy})\phi(\bfy)\,,
\label{eq28}\\
H_{\rm int}^{(8)}&=&-\frac{|e|Z}{M}
\int d\bfy\;\bfA(\bfy)
\phi^{\dag}(\bfy)(-i\bnabla_{\bfy})\phi(\bfy)\,,
\label{eq29}\\
H_{\rm int}^{(9)}&=&-e\int d\bfx\;
\psi^{\dag}(\bfx)\balpha \cdot \bfA_{\rm cl}(\bfx)\psi(\bfx)\,,
\label{eq30}\\
H_{\rm int}^{(10)}&=&\frac{e}{2}\frac{m}{M}\int d\bfx\;
\psi^{\dag}(\bfx)\balpha\psi(\bfx)
\int d\bfy\;
\psi^{\dag}(\bfy)\bfA_{\rm cl}(\bfy)\psi(\bfy)\nonumber\\
&&+\frac{e}{2}\frac{m}{M}\int d\bfy\;
\psi^{\dag}(\bfy)\bfA_{\rm cl}({\bfy})\psi(\bfy)
\int d\bfx\;
\psi^{\dag}(\bfx)\balpha\psi(\bfx)\,,
\label{eq31}\\
H_{\rm int}^{(11)}&=&-e\int d\bfx\;
\psi^{\dag}(\bfx)\balpha\psi(\bfx)
\int d\bfy\;
\phi^{\dag}(\bfy)\bfA_{\rm cl}({\bfy})\phi(\bfy)\,,
\label{eq32}\\
H_{\rm int}^{(12)}&=&-\frac{|e|Z}{M}
\int d\bfy\;\bfA_{\rm cl}(\bfy)
\phi^{\dag}(\bfy)(-i\bnabla_{\bfy})\phi(\bfy)\,,
\label{eq33}\\
H_{\rm int}^{(13)}&=&\frac{|e|Z}{M}\int d\bfx\;
\psi^{\dag}(\bfx)(-i\bnabla_{\bfx})\psi(\bfx)
\int d\bfy\;
\phi^{\dag}(\bfy)\bfA_{\rm cl}({\bfy})\phi(\bfy)\,.
\label{eq34}
\ee
Here $\psi(\bfx)$ and $\phi(\bfx)$ are
 the electron and nucleus field operators in the Schr\"odinger 
representation and $\Delta_{\bfx}$ is the Laplace operator.
The scalar product is implicit in equations 
(\ref{eq23})-(\ref{eq34}). It should be stressed that, despite
the nuclear field operator $\phi(\bfx)$ is introduced,
the Hamiltonian (\ref{eq20}) has physical sense only in the
one nuclear particle sector. The introduction of $\phi(\bfx)$
will simplify the use of the standard QED methods for calculations
by perturbation theory.

The total momentum operator of the system is given by
\be \label{eq35}
{\bf {\cal P}}=\int d\bfx\; \phi^{\dag}(-i\bnabla_{\bfx})\phi(\bfx)
+\int d{\bf x}\;
[{\bf {\cal E}}_{t}({\bf x})\times{\bf {\cal H}}({\bf x})]\,.
\ee
This operator commutes with the Hamiltonian of the system
if the external magnetic field is switched off, 
$\bfA_{\rm cl}(\bfx)=0$. It also commutes with the
operator
\be \label{eq36}
H'=H_0+\sum_{s=1}^{s=10}H_{\rm int}^{(s)}\,.
\ee
In the theory without the terms $H_{\rm int}^{(11)}$,
$H_{\rm int}^{(12)}$, and $H_{\rm int}^{(13)}$ as well as
in the theory without the external magnetic field at all,
operators $\phi(\bfx)$, $\bfA(\bfx)$, and
 $\psi(\bfx)$ obey the following space-shift 
transformation rules
\be \label{eq37}
\phi(\bfx+\bfa)&=&\exp{(-i\bfa\cdot{\bf {\cal P}})}\phi(\bfx)
\exp{(i\bfa\cdot{\bf {\cal P}})}\,,\\
\bfA(\bfx+\bfa)&=&\exp{(-i\bfa\cdot{\bf {\cal P}})}\bfA(\bfx)
\exp{(i\bfa\cdot{\bf {\cal P}})}\,,
\label{eq37a}
\\
\psi(\bfx)&=&\exp{(-i\bfa\cdot{\bf {\cal P}})}\psi(\bfx)
\exp{(i\bfa\cdot{\bf {\cal P}})}\,.
\label{eq37b}
\ee
For any operator $F(\bfx)$, 
the transition to the Heisenberg representation
is performed by the substitution
\be \label{eq38}
F_{\rm S}(\bfx)=\exp{(-iHt)}F_{\rm H}(t,\bfx)
\exp{(iHt)}\,.
\ee

The calculations of the atomic energy levels,
based on the Hamiltonian (\ref{eq20}), can be performed
by perturbation theory. It is convenient to consider
that in zeroth approximation the Hamiltonian of the system
is given by the term $H_0$. The terms 
$H_{\rm int}^{(1)}$ - $H_{\rm int}^{(13)}$
are accounted for by perturbation theory.
As usual (see, e.g., \cite{dav63,bjo65}),
the vacuum state is defined as a state in which
the negative-energy continuum is occupied by electrons and the
electron current operator $e\overline{\psi}(x)\gamma_{\mu}\psi(x)$
must be replaced by  $(e/2)[\overline{\psi}(x)\gamma_{\mu},\psi(x)]$
in all the equations given above.

\begin{center}
{\bf III.$\;$ GENERAL FORMULAS FOR THE ENERGY SHIFTS}
\end{center}

First, we formulate a procedure for deriving the
energy shift due to the terms $H_{\rm int}^{(1)}$ -
 $H_{\rm int}^{(10)}$. The operator $H'$ defined
by equation (\ref{eq36}) commutes with the total momentum
of the system (\ref{eq35}) and, therefore, the atom
can be characterized by a fixed momentum $\bfK$.
We are interested in the energy shift,
$\Delta E_{(a,\bfK)}=E_{(a,\bfK)}-E_{(a,\bfK)}^{(0)}$, 
of a single isolated level 
$|a,\bfK\rangle$ of an $N$-electron atom. Here
$N$ denotes the number of the atomic (positive-energy-state)
electrons, $\bfK$ is the total momentum of the system
described by the Hamiltonian $H'$, and $a$ denotes 
the set of the other quantum numbers.
In zeroth approximation, the energy $E_{(a,\bfK)}^{(0)}$ is
\be \label{eq39}
E_{(a,\bfK)}^{(0)}=\veps_{a_1}+\cdots \veps_{a_N}+\frac{\bfK^2}{2M}\,.
\ee
The one-electron energies $\veps_n$ are defined by the Dirac
equation
\be \label{eq40}
(-i\balpha\cdot\bnabla+\beta m+V(\bfx))\psi_n(\bfx)=
\veps_n\psi_n(\bfx)\,, 
\ee
where $V(\bfx)$ is the Coulomb potential of the nucleus.
We note that the formalism considered here
allows us to account partially for the finite nuclear size 
effect by employing in (\ref{eq40})
 the potential of an extended nucleus.
We could start also with the Dirac equation with an effective
potential $V_{\rm eff}(\bfx)$ (e.g., a local version of the
Hartree-Fock potential) which approximately describes
the electron-electron interaction. In this case the interaction
with the potential $\Delta V(\bfx)=V(\bfx)-V_{\rm eff}(\bfx)$
must be accounted for perturbatively to eliminate double accounting
the interelectronic interaction corrections.
The electronic part of the unperturbed wave function
is a linear combination of one-determinant functions
\be \label{eq41}
u_a(\bfx_1,...,\bfx_N)=\sum_{b}C_a^b\frac{1}{\sqrt{N!}}
\sum_{P}(-1)^P\psi_{Pb_1}(\bfx_1)\cdots \psi_{Pb_N}(\bfx_N)\,.
\ee
We introduce the Green function $g_a(E,\bfK)$ by
\be \label{eq42}
g_a(E,\bfK)\delta(E'-E)\delta(\bfK'-\bfK)&=&
\frac{1}{2\pi i}\frac{1}{N!}\frac{1}{(2\pi)^3}
\int d\bfx_1\cdots d\bfx_N d\bfx'_1\cdots d\bfx'_N d\bfX\, d\bfX'
\nonumber\\
&&\times \int_{-\infty}^{\infty}dt\, dt'\;\exp{(iE't'-iEt)}
\exp{(-i\bfK'\cdot\bfX'+i\bfK\cdot\bfX)}\nonumber\\
&&\times u_a^{\dag}(\bfx'_1,...,\bfx'_N)
 \la 0|T\psi(t',\bfx'_1)\cdots\psi(t',\bfx'_N)
\phi(t',\bfX')\nonumber\\
&&\times\phi^{\dag}(t,\bfX)\psi^{\dag}(t,\bfx_N)
\cdots \psi^{\dag}(t,\bfx_1)|0\ra
 u_a(\bfx_1,...,\bfx_N)\,,
\ee
where $\psi(t,\bfx)$ and $\phi(t,\bfx)$ are the electron
and nucleus field operators in the Heisenberg representation
and $T$ is the time-ordering operator.
As one can see from equation (\ref{eq42}), 
$g_a(E,\bfK)$ is the Fourier transform of the two-time
Green function. The $g_a(E,\bfK)$ function contains the complete 
information about the energy levels of the system
described by $H'$.
Defined by equation (\ref{eq42}) for real $E$, it can be continued 
analytically to  the complex $E$ plane with some cuts
(see, e.g., \cite{log63,sha01}).
From the spectral representation of $g_a(E,\bfK)$,
we obtain
\be \label{eq43}
g_a(E,\bfK)=\frac{\varphi_{(a,\bfK)}\varphi_{(a,\bfK)}^*}
{E-E_{(a,\bfK)}}+\mbox{ terms that are regular at }
E\sim E_{(a,\bfK)}\,,
\ee
where
\be \label{eq44}
\varphi_{(a,\bfK)}=\frac{(2\pi)^{3/2}}{\sqrt{N!}}\int d\bfx_1
\cdots d\bfx_N\; u_a^{\dag}(\bfx_1,...,\bfx_N)
\la 0|\psi(0,\bfx_1)\cdots\psi(0,\bfx_N)
\phi(0,0)|a,\bfK\ra \,.
\ee
As in \cite{sha01}, using equation (\ref{eq43}),
we can derive the following formula for the energy
shift
\be \label{eq45}
  \Delta E_{(a,\bfK)} = \frac{\displaystyle
\frac{1}{2\pi i}
        \oint_{\Gamma} dE\;
(E-E_{(a,\bfK)}^{(0)}) \Delta g_{a}(E,\bfK)}
      {\displaystyle
1+\frac{1}{2\pi i}\oint_{\Gamma}
 dE \;\Delta g_{a}(E,\bfK)}\,,  
\ee
where  $\Delta g_{a}(E,\bfK)=g_{a}(E,\bfK)- g_{a}^{(0)}(E,\bfK)$
and $g_{a}^{(0)}(E,\bfK)=(E-E_{(a,\bfK)}^{(0)})^{-1}$.
The contour $\Gamma$
 surrounds the pole corresponding to the level 
$a$ and keeps outside all other singularities of 
$g_{a}(E,\bfK)$. It is assumed that the contour $\Gamma$
is oriented  anticklockwise. The Green function $g_{a}(E,\bfK)$
is constructed by perturbation theory after the transition
in equation (\ref{eq42}) to the interaction representation.
Since we are interested in the energy shifts in the
atom rest system, we must put $\bfK=0$ in formula (\ref{eq45}).

Formula (\ref{eq45}) can be used to calculate the corrections
due to the terms $H_{\rm int}^{(1)}$-$H_{\rm int}^{(10)}$.
The term $\Delta H'$, defined as
\be \label{eq46}
\Delta H'=H_{\rm int}^{(11)}+H_{\rm int}^{(12)}+H_{\rm int}^{(13)}\,,
\ee
does not commute with ${\bf {\cal P}}$ and, therefore, 
requires a special treatment. To first order in the
interaction with the magnetic field, the energy shift
due to this term can be written as (cf., \cite{fau70c})
\be \label{eq47}
\Delta E_{(a,\bfK)}\delta(\bfK'-\bfK)\simeq 
\la a,\bfK'|\Delta H'|a,\bfK\ra\,,
\ee
where $|a,\bfK\ra$ and $|a,\bfK'\ra$ are the eigenvectors
of the operator $H'$. Since we are interested in the corrections
of first order in the magnetic field, in equation (\ref{eq47})
 we can consider  
the eigenvectors of the operator
 $H''=H'- H_{\rm int}^{(9)} -H_{\rm int}^{(10)}$.
We assume that they are normilized by
\be \label{eq48}
\la a,\bfK'|a,\bfK\ra=\delta(\bfK'-\bfK)\,.
\ee
We introduce the Green function 
 $g_a(E',E,\bfK',\bfK)$ by
\be \label{eq49}
g_a(E',E,\bfK',\bfK)&=&
-\frac{1}{N!}\frac{1}{(2\pi)^3}
\int d\bfx_1\cdots d\bfx_N d\bfx'_1\cdots d\bfx'_N d\bfX\, d\bfX'
\nonumber\\
&&\times \int_{-\infty}^{\infty}dt\, dt'\;\exp{(iE't'-iEt)}
\exp{(-i\bfK'\cdot\bfX'+i\bfK\cdot\bfX)}\nonumber\\
&&\times u_a^{\dag}(\bfx'_1,...,\bfx'_N)
 \la 0|T\psi(t',\bfx'_1)\cdots\psi(t',\bfx'_N)
\phi(t',\bfX') \Delta H'(0)\nonumber\\
&&\times\phi^{\dag}(t,\bfX)\psi^{\dag}(t,\bfx_N)
\cdots \psi^{\dag}(t,\bfx_1)|0\ra
 u_a(\bfx_1,...,\bfx_N)\,,
\ee
where, as in equation (\ref{eq42}), the Heisenberg
representation is used and $\Delta H'(0)\equiv \Delta H'(t=0)$.
The analytical properties of this type Green function
as a function of two complex variables $E$ and $E'$
in the region $E\sim E_{(a,\bfK)}$,
$E'\sim E_{(a,\bfK')}$
can be investigated by deriving the double spectral
representation (see, e.g., \cite{sha01,fau70c}).
From this representation, we obtain
\be \label{eq50}
g_a(E',E,\bfK',\bfK)&=&
\frac{\varphi_{(a,\bfK')}}{E'-E_{(a,\bfK')}}
\la a,\bfK'|\Delta H'(0)|a,\bfK\ra
\frac{\varphi_{(a,\bfK)}^*}{E-E_{(a,\bfK)}}\nonumber\\
&&+ \mbox{ terms that are regular functions of } E'
\mbox{ or } E \mbox{ when } \nonumber\\
&& E'\sim E_{(a,\bfK')}
\mbox{ and} E\sim E_{(a,\bfK)}\,.
\ee
As in the case of the photon emission by an atom
considered in detail in \cite{sha01}, from equations
(\ref{eq47}), (\ref{eq50}) we derive
\be \label{eq51}
  \Delta E_{(a,\bfK)}\delta(\bfK'-\bfK)
 \simeq \frac{\displaystyle
\frac{1}{2\pi i}
        \oint_{\Gamma} dE\;
\frac{1}{2\pi i}
        \oint_{\Gamma'} dE'\;
 g_{a}(E',E,\bfK',\bfK)}
      {\displaystyle
\Bigl[\frac{1}{2\pi i}\oint_{\Gamma'} dE\;
g_{a}(E,\bfK')\Bigr]^{1/2}
\Bigl[\frac{1}{2\pi i}\oint_{\Gamma} dE\;
g_{a}(E,\bfK)\Bigr]^{1/2}}\,,  
\ee
where the contours $\Gamma$ and
$\Gamma'$ surround the poles corresponding to the levels
$E_{(a,\bfK)}$ and $E_{(a,\bfK')}$, respectively,
and keep outside all other singularities.
 The Green function
$g_{a}(E',E,\bfK',\bfK)$ is constructed by perturbation theory
after  the transition in (\ref{eq49}) to the interaction
representation. Since we are interested in the energy
shifts in the atom rest system, at the end of the calculation
by formula (\ref{eq51}), we must put
$\bfK=0$.

\begin{center}
{\bf IV.$\;$RELATIVISTIC FORMULA FOR THE RECOIL CORRECTION
TO THE BOUND-ELECTRON $g$ FACTOR IN A HYDROGENLIKE ATOM} 
\end{center}

The method formulated in the previous sections can be used to
derive the complete $\alpha Z$-dependence expressions for
the recoil corrections  to the energy levels 
($\bfA_{\rm cl}(\bfx)=0$) and to the 
atomic $g$ factor to first order in $m/M$. The relevant
expression for the recoil correction to the energy levels
in a hydrogenlike atom was first derived by a quasipotential
method in \cite{sha85} and was subsequently rederived by
other methods in \cite{yel94,pac95,sha98b}
(a part of the expression was previously found in \cite{bra73}).
In this section, we derive the corresponding formula
for the recoil correction to the bound-electron $g$ factor
in a hydrogenlike atom. 

In what follows, we will consider that $V(\bfx)$ in the
Dirac equation (\ref{eq40}) is 
the potential of an extended-charge nucleus. This will
allow us to avoid a special treatment of the degenerate
states of different parity (such as $2s$ and $2p_{1/2}$)
in the derivation of the contributions from the $\Delta H'$
term. However, as it will be seen from
 the final formulas obtained below, they have neither 
singularities nor ambiguities when
$V(\bfx)\rightarrow V_{\rm C}(\bfx)=-\alpha Z/|\bfx|$
and, therefore, the pure Coulomb potential can be used
in these formulas as well. 

We are interested in the energy shifts of first order in
the magnetic field and of first order in $m/M$. The contributions
of interest arise in a few lowest orders of the perturbation theory.
They can be conventionally represented by the following combinations:
\be \label{eq52}
\Delta E^{(1)}& \sim & H_{\rm int}^{(9)}\times H_{\rm int}^{(3)}
+H_{\rm int}^{(9)}\times H_{\rm int}^{(4)}\times H_{\rm int}^{(2)}
\nonumber\\
&&+H_{\rm int}^{(9)}\times
H_{\rm int}^{(5)}\times H_{\rm int}^{(2)}\times H_{\rm int}^{(2)}
+ \mbox{ permutations}\,,\\
\Delta E^{(2a)} & \sim &H_{\rm int}^{(10)}\,,
\label{eq53} \\
\Delta E^{(2b)} & \sim &H_{\rm int}^{(11)}\times H_{\rm int}^{(6)}
+H_{\rm int}^{(11)}\times H_{\rm int}^{(7)}\nonumber\\
&&+H_{\rm int}^{(11)}\times H_{\rm int}^{(8)}\times
H_{\rm int}^{(2)}+\mbox{ permutations}\,.
\label{eq54}
\ee
As to the terms $H_{\rm int}^{(12)}$ and 
$H_{\rm int}^{(13)}$, in the order under 
consideration, the first of
these terms gives a contribution which is independent
of the internal atomic quantum numbers while the second one
gives zero when it is averaged with the electron wave function
of a definite parity. Therefore, the both terms can be omitted.

To calculate $\Delta E^{(1)}$, it is convenient to consider
that $H_{\rm int}^{(9)}$ is included in the unperturbed 
Hamiltonian. This means that the interaction with the
magnetic field, $\delta V(\bfx)=-e\balpha \cdot \bfA_{\rm cl}(\bfx)$,
must be included in the Dirac equation (\ref{eq40}).
Then the derivation of the formula for this correction
reduces to the related derivation of the recoil 
correction to the energy levels for 
$\bfA_{\rm cl}(\bfx)=0$ \cite{sha85,yel94,pac95,sha98b}.
 It follows that to obtain
$\Delta E^{(1)}$, we may simply replace $V(\bfx)$ by
$V(\bfx)+\delta V(\bfx)$ everywhere
in the formula for the recoil correction to the energy level
and decompose it to first order in $\delta V(\bfx)$. 
However, before to do that, let us demonstrate how the
recoil correction to the energy level
can be derived within the formalism
considered here.

Let us consider, for example, the contribution
$\sim H_{\rm int}^{(4)}\times 
H_{\rm int}^{(2)}+
H_{\rm int}^{(2)}\times 
H_{\rm int}^{(4)}$.
In the interaction representation, the related
contribution to the two-time Green function is
\be \label{eq55}
\Delta G(t',\bfx',\bfX';t,\bfx,\bfX)&=&
\frac{(-i)^2}{2}\int_{-\infty}^{\infty}dt_1 dt_2\;
\la 0|T\psi(t',\bfx')\phi(t',\bfX')\psi^{\dag}(t,\bfx)
\phi^{\dag}(t,\bfX)\nonumber\\
&&\times \Bigl\{\frac{(-e)}{2}\int d\bfx_1 d\bfX_1\;
[\psi^{\dag}(t_1,\bfx_1)\alpha^{i},\psi(t_1,\bfx_1)]
A^i(t_1,\bfx_1+\bfX_1)\nonumber\\
&&\times \phi^{\dag}(t_1,\bfX_1)
\phi(t_1,\bfX_1) \frac{|e|Z}{M}
\int d\bfy_1 d\bfY_1\; 
\psi^{\dag}(t_2,\bfy_1)\nonumber\\
&&\times (-i\nabla^{k}_{\bfy_1})\psi(t_2,\bfy_1)
A^k(t_2,\bfY_1)\phi^{\dag}(t_2,\bfY_1)
\phi(t_2,\bfY_1)\nonumber\\
&&+  \frac{|e|Z}{M}
\int d\bfy_1 d\bfY_1\; 
\psi^{\dag}(t_2,\bfy_1)(-i\nabla^{k}_{\bfy_1})\psi(t_2,\bfy_1)
A^k(t_2,\bfY_1)\nonumber\\
&&\times \phi^{\dag}(t_2,\bfY_1)
\phi(t_2,\bfY_1)\frac{(-e)}{2}\int d\bfx_1 d\bfX_1\;
[\psi^{\dag}(t_1,\bfx_1)
\alpha^{i},\psi(t_1,\bfx_1)]\nonumber\\
&&\times A^i(t_1,\bfx_1+\bfX_1) \phi^{\dag}(t_1,\bfX_1)
\phi(t_1,\bfX_1)\Bigr\}|0\ra_{\rm con} \,,
\ee
where all the operators are considered in the interaction
representation and the summation over the repeated indices
$(i,k=1,2,3)$, which enumerate components of the three-dimensional
vectors, is implicit. The label "con" means that
contributions containing
disconnected vacuum-vacuum terms must be omitted.
According to the permutation
rules for the $T$ product of the boson and fermion operators,
 second term in the right hand side of equation
(\ref{eq55}) is equal to first one. 
Assuming that the time-ordering operator is defined for
equal-time fermion operators by \cite{moh98}
\be \label{timeord}
T[A(t)B(t)]=\frac{1}{2}A(t)B(t)-\frac{1}{2}B(t)A(t)\,,
\ee
we have
\be \label{eq56}
\Delta G(t',\bfx',\bfX';t,\bfx,\bfX)&=&
\frac{e|e|Z}{M}\int_{-\infty}^{\infty}dt_1 dt_2\;
\la 0|T\psi(t',\bfx')\phi(t',\bfX')\psi^{\dag}(t,\bfx)
\phi^{\dag}(t,\bfX)\nonumber\\
&&\times \int d\bfx_1 d\bfX_1\;
\psi^{\dag}(t_1,\bfx_1)\alpha^{i}\psi(t_1,\bfx_1)
A^i(t_1,\bfx_1+\bfX_1)\nonumber\\
&&\times \phi^{\dag}(t_1,\bfX_1)
\phi(t_1,\bfX_1) 
\int d\bfy_1 d\bfY_1\; 
\psi^{\dag}(t_2,\bfy_1)\nonumber\\
&&\times (-i\nabla^{k}_{\bfy_1})\psi(t_2,\bfy_1)
A^k(t_2,\bfY_1)\phi^{\dag}(t_2,\bfY_1)
\phi(t_2,\bfY_1)|0\ra_{\rm con}\,.
\ee
Using the Wick theorem and keeping only the terms which 
will contribute to the state-dependent energy shift, we obtain
\be \label{eq57}
\Delta G(t',\bfx',\bfX';t,\bfx,\bfX)&=&
\frac{e|e|Z}{M}\int_{-\infty}^{\infty}dt_1 dt_2\;
\int d\bfx_1 d\bfX_1 d\bfy_1 d\bfY_1 
[S(t'-t_1,\bfx',\bfx_1)\nonumber\\
&&\times \alpha^i S(t_1-t_2,\bfx_1,\bfy_1)
(-i\nabla^k_{\bfy_1})S(t_2-t,\bfy_1,\bfx)\nonumber\\
&&+S(t'-t_2,\bfx',\bfy_1)(-i\nabla^k_{\bfy_1})
S(t_2-t_1,\bfy_1,\bfx_1)\nonumber\\
&&\times \alpha^iS(t_1-t,\bfx_1,\bfx)]
 D^{ik}(t_1-t_2,\bfx_1+\bfX_1-\bfY_1)\nonumber\\
&&\times [S_n(t'-t_1,\bfX'-\bfX_1)S_n(t_1-t_2,\bfX_1-\bfY_1)
\nonumber\\
&&\times S_n(t_2-t,\bfY_1-\bfX)+
S_n(t'-t_2,\bfX'-\bfY_1)\nonumber\\
&&\times S_n(t_2-t_1,\bfY_1-\bfX_1)
S_n(t_1-t,\bfX_1-\bfX)]\,.
\ee
Here 
\be \label{eq58}
S(t'-t,\bfx,\bfy)=\frac{i}{2\pi}\int_{-\infty}^{\infty}
d\omega \; \exp{[-i\omega(t'-t)]}
\sum_n\frac{\psi_n(\bfx)\psi_n^{\dag}(\bfy)}
{\omega-\veps_n(1-i0)}
\ee
is the electron propagator in the nuclear potential $V(\bfx)$,
\be \label{eq59}
D^{il}(t'-t,\bfx-\bfy)=
\frac{i}{2\pi}\int_{-\infty}^{\infty}
d\omega \; \exp{[-i\omega(t'-t)]}
\int \frac{d\bfk}{(2\pi)^3}\;
\frac{\exp{[i\bfk\cdot(\bfx-\bfy)]}}
{\omega^2-\bfk^2+i0}\Bigl(\delta_{il}-\frac{k^i k^l}{\bfk^2}
\Bigr)
\ee
is the transverse part of the photon propagator in the
Coulomb gauge, and
\be \label{eq60}
S_n(t'-t,\bfX-\bfY)=
\frac{i}{2\pi}\int_{-\infty}^{\infty}
d\omega \; \exp{[-i\omega(t'-t)]}  \int \frac{d\bfK}{(2\pi)^3}\;
\frac{\exp{[i\bfK\cdot(\bfX-\bfY)]}}
{\omega-\frac{\bfK^2}{2M}+i0}
\ee
is the nucleus propagator. In equation (\ref{eq58}) the index
$n$ runs over all bound and continuum states.
 Since we are interested in the
corrections of first order in $m/M$ and we have already 
the factor $1/M$ in front of expression (\ref{eq57}),
we can consider the limit $M\rightarrow 0$ in the expression
for the nucleus propagator. In this limit, we have 
\be \label{eq61}
S_n(t'-t,\bfX-\bfY)=\theta(t'-t)\delta(\bfX-\bfY)\,.
\ee
Using this expression for $S_n$, we obtain
\be \label{eq62}
\Delta G(t',\bfx',\bfX';t,\bfx,\bfX)&=&
\frac{e|e|Z}{M}\int_{-\infty}^{\infty}dt_1 dt_2\;
\int d\bfx_1 d\bfy_1\;  
[S(t'-t_1,\bfx',\bfx_1)\nonumber\\
&&\times \alpha^i S(t_1-t_2,\bfx_1,\bfy_1)
(-i\nabla^k_{\bfy_1})S(t_2-t,\bfy_1,\bfx)\nonumber\\
&&+S(t'-t_2,\bfx',\bfy_1)(-i\nabla^k_{\bfy_1})
S(t_2-t_1,\bfy_1,\bfx_1)\nonumber\\
&&\times \alpha^iS(t_1-t,\bfx_1,\bfx)]
 D^{ik}(t_1-t_2,\bfx_1)\nonumber\\
&&\times \delta(\bfX'-\bfX)
[\theta(t'-t_1)\theta(t_1-t_2)
\theta(t_2-t)\nonumber\\
&&+\theta(t'-t_2)
\theta(t_2-t_1)
\theta(t_1-t)]\,.
\ee
The related contribution to the Green function
$g_a(E,\bfK)$ defined by equation (\ref{eq42}) is
\be \label{eq63}
\Delta g_a(E,\bfK)\delta(E'-E)\delta(\bfK'-\bfK)&=&
\frac{1}{2\pi i} \frac{1}{(2\pi)^3}
\frac{e|e|Z}{M}\int_{-\infty}^{\infty} dt\,dt'\,
dt_1\, dt_2\;
\int d\bfx\, d\bfx'\, d\bfX\, d\bfX'\nonumber\\
&&\times  d\bfx_1\, d\bfy_1\;\exp{(iE't'-iEt)}\nonumber\\
&&\times \exp{(-i\bfK'\cdot \bfX'+i\bfK\cdot \bfX)}
\psi_a^{\dag}(\bfx')
[S(t'-t_1,\bfx',\bfx_1)\nonumber\\
&&\times \alpha^i S(t_1-t_2,\bfx_1,\bfy_1)
(-i\nabla^k_{\bfy_1})S(t_2-t,\bfy_1,\bfx)\nonumber\\
&&+S(t'-t_2,\bfx',\bfy_1)(-i\nabla^k_{\bfy_1})
S(t_2-t_1,\bfy_1,\bfx_1)\nonumber\\
&&\times \alpha^iS(t_1-t,\bfx_1,\bfx)]\psi_a(\bfx)
D^{ik}(t_1-t_2,\bfx_1)\nonumber\\
&&\times \delta(\bfX'-\bfX)
[\theta(t'-t_1)\theta(t_1-t_2)
\theta(t_2-t)\nonumber\\
&&+\theta(t'-t_2)
\theta(t_2-t_1)
\theta(t_1-t)]\,.
\ee
Integrating over $\bfX$, $\bfX'$, and over the time
variables and setting $\bfK=0$, we obtain
\be \label{eq64}
\Delta g_a(E,0)&=&-\frac{e|e|Z}{M}
\frac{1}{(E-\veps_a)^2}\int_{-\infty}^{\infty}d\omega\;
\Bigl\{\sum_n\frac{\la a|\alpha^iD^{ik}(\omega)|n\ra
\la n|(-i\nabla^k)|a\ra}{E-\omega-\veps_n(1-i0)}
\nonumber\\
&&+
\sum_n\frac{\la a|(-i\nabla^k)|n\ra 
\la n|\alpha^iD^{ik}(\omega)|a\ra}{E-\omega-\veps_n(1-i0)}
\Bigr\}\,,
\ee
where
\be \label{eq65}
D^{il}(\omega,{\bf r})&=&
\int \frac{d\bfk}{(2\pi)^3}\;
\frac{\exp{(i\bfk\cdot{\bf r})}}
{\omega^2-\bfk^2+i0}\Bigl(\delta_{il}-\frac{k^i k^l}{\bfk^2}
\Bigr)\nonumber\\
&=&-\frac{1}{4\pi}\Bigl\{\frac
{\exp{(i|\omega|r)}}{r}\delta_{il}+\nabla^{i}\nabla^{l}
\frac{(\exp{(i|\omega|r)}
-1)}{\omega^{2}r}\Bigr\}\,.
\ee
The corresponding contribution to the energy shift
according to equation (\ref{eq45}) for $\bfK=0$ is
\begin{eqnarray} \label{eq66}
\Delta E&=&\frac{4\pi\alpha Z}{M}\frac{i}{2\pi}
\int_{-\infty}^{\infty}d\omega\,\sum_{n}\Biggl\{
\frac{
\langle a|
\alpha^{i} D^{ik}(\omega)|n\rangle
\langle n|(-i\nabla^{k})|a\rangle}
{\veps_a-\omega-\varepsilon_{n}(1-i0)}\nonumber\\
&&+\frac{
\langle a|(-i\nabla^{k})|n\rangle\langle n|
\alpha^{i} D^{ik}(\omega)|a\rangle}
{\veps_a-\omega-\varepsilon_{n}(1-i0)}\Biggr\}\,.
\end{eqnarray}

The energy shifts due to the terms $H_{\rm int}^{(3)}$
and $H_{\rm int}^{(5)}$
($\sim H_{\rm int}^{(5)}\times H_{\rm int}^{(2)}
\times H_{\rm int}^{(2)}$ + permutations)
 can be derived in the same way.
The total contribution of the terms $H_{\rm int}^{(3)}$,
$H_{\rm int}^{(4)}$
($\sim H_{\rm int}^{(4)}\times H_{\rm int}^{(2)}
+H_{\rm int}^{(2)}\times H_{\rm int}^{(4)}$),
 and $H_{\rm int}^{(5)}$ 
($\sim H_{\rm int}^{(5)}\times H_{\rm int}^{(2)}
\times H_{\rm int}^{(2)}$ + permutations) is
\begin{eqnarray} \label{eq67}
\Delta E=\frac{i}{2\pi M}
\int_{-\infty}^{\infty}d\omega\;
\langle a|[p^k- D^{k}(\omega)]
G(\omega+\veps_a)[p^k- D^{k}(\omega)]|a\rangle\,,
\end{eqnarray}
where $p^k=-i\nabla^k$, 
\be \label{eq68}
D^k(\omega)=-4\pi\alpha Z\alpha^l D^{lk}(\omega)\,,
\ee
and
\be \label{greencoul}
G(\omega)=\sum_n\frac{|n\ra \la n|}{\omega-\veps_n(1-i0)}\,.
\ee
Deriving equation (\ref{eq67}) we considered that in the
zeroth approximation the electron obeys the Dirac equation
with the nuclear potential $V(\bfx)$. As was indicated above,
to obtain the desired formula for $\Delta E^{(1)}$, defined
by equation (\ref{eq52}), we should replace $V(\bfx)$
by $V(\bfx)+\delta V(\bfx)$,  where
$\delta V(\bfx)=-e\balpha \cdot\bfA_{\rm cl}(\bfx)$,
in equation (\ref{eq67}) and expand it to first order
in $\delta V(\bfx)$. As a result of this expansion,
we obtain 
\begin{eqnarray} \label{eq69}
\Delta E^{(1)}&=&\frac{i}{2\pi M}
\int_{-\infty}^{\infty}d\omega\;
\{\langle \delta a|[p^k- D^{k}(\omega)]
G(\omega+\veps_a)[p^k- D^{k}(\omega)]|a\rangle\nonumber\\
&&+\langle a|[p^k- D^{k}(\omega)]
G(\omega+\veps_a)[p^k- D^{k}(\omega)]|\delta a\rangle\nonumber\\
&&+\langle a|[p^k- D^{k}(\omega)]
G(\omega+\veps_a)(\delta V-\delta \veps_a)\nonumber\\
&&\times G(\omega+\veps_a)[p^k- D^{k}(\omega)]|a\rangle\}\,,
\end{eqnarray}
where
\be \label{eq70}
\delta \veps_a&=&\la a|\delta V|a\ra\,,\\
|\delta a\ra&=&\sum_n^{\veps_n\ne \veps_a}\frac{|n\ra\la n|\delta V|a\ra}
{\veps_a-\veps_n}\,.
\label{eq71}
\ee
For practical calculations, it is convenient
to represent expression (\ref{eq69})
by the sum of the lower-order term and the higher-order term,
\be \label{eq72}
\Delta E^{(1)}&=&\Delta E^{(1)}_{\rm L}+\Delta E^{(1)}_{\rm H}\,,
\\
\Delta E^{(1)}_{\rm L}&=&\frac{1}{2M}\{\la \delta a|
[p^k-2D^k(0)]p^k|a\ra
+\la a|[p^k-2D^k(0)]p^k|\delta a\ra\} \,,
\label{eq73}
\\
\Delta E^{(1)}_{\rm H}&=&\frac{i}{2\pi M}
\int_{-\infty}^{\infty} d\omega\;\Bigl\{
\la \delta a|\Bigl(D^k(\omega)-\frac{[p^k,V]}{\omega+i0}\Bigr)
G(\omega+\veps_a)\Bigl(D^k(\omega)+\frac{[p^k,V]}{\omega+i0}\Bigr)|a\ra
\nonumber\\
&&+\la a|\Bigl(D^k(\omega)-\frac{[p^k,V]}{\omega+i0}\Bigr)
G(\omega+\veps_a)\Bigl(D^k(\omega)+\frac{[p^k,V]}{\omega+i0}\Bigr)
|\delta a\ra\nonumber\\
&&+ \la a|\Bigl(D^k(\omega)-\frac{[p^k,V]}{\omega+i0}\Bigr)
G(\omega+\veps_a)(\delta V-\delta \veps_a)G(\omega+\veps_a)\nonumber\\
&&\times \Bigl(D^k(\omega)+\frac{[p^k,V]}{\omega+i0}\Bigr)|a\ra
\nonumber\\
&&-\la a|\frac{[p^k,\delta V]}{\omega+i0}
G(\omega+\veps_a)\Bigl(D^k(\omega)+\frac{[p^k,V]}{\omega+i0}\Bigr)|a\ra
\nonumber\\
&&+\la a|\Bigl(D^k(\omega)-\frac{[p^k,V]}{\omega+i0}\Bigr)
G(\omega+\veps_a)\frac{[p^k,\delta V]}{\omega+i0}|a\ra\Bigr\}\,,
\label{eq74}
\ee
where $[A,B]=AB-BA$.
The term $\Delta E^{(1)}_{\rm L}$ contains the non-relativistic
and lowest-order relativistic contributions and a part of
 the higher-order relativistic contributions.
 The term $\Delta E^{(1)}_{\rm H}$
contains all the higher-order relativistic contributions
 which are not included into
$\Delta E^{(1)}_{\rm L}$.

The direct evaluation of the term $H_{\rm int}^{(10)}$
yields
\be \label{eq75}
\Delta E^{(2a)}&=&e\frac{m}{M}\frac{i}{2\pi}
\int_{-\infty}^{\infty}d\omega\;
[\langle a|\alpha^{k}G(\omega)A_{\rm cl}^{k}|a\rangle 
+\langle a|A_{\rm cl}^{k}G(\omega)\alpha^{k}|a\rangle]\,.
\ee
The calculation of the correction $\Delta E^{(2b)}$
is a more difficult task. Here
we derive in detail  the contribution which
appears as a combination $ \sim\; H_{\rm int}^{(11)}
\times H_{\rm int}^{(8)}\times H_{\rm int}^{(2)}$
+ permutations. To find this correction we should
use formula (\ref{eq51}). The related contribution
to the Green function $g_a(E',E,\bfK',\bfK)$
defined by equation (\ref{eq49}) is
\be \label{eq76}
\Delta g_a(E',E,\bfK',\bfK)&=&
\frac{1}{(2\pi)^3}(-i)^2
\frac{e^2|e|Z}{M}\int_{-\infty}^{\infty} dt\,dt'\,
dt_1\, dt_2\;
\int d\bfx\, d\bfx'\, d\bfX\, d\bfX'\nonumber\\
&&\times \exp{(iE't'-iEt)}
\exp{(-i\bfK'\cdot \bfX'+i\bfK\cdot \bfX)}
\psi_a^{\dag}(\bfx')\nonumber\\
&&\times \la 0|T\psi(t',\bfx')\phi(t',\bfX')
\int d\bfz_1 d\bfZ_1 \psi^{\dag}(0,\bfz_1)
\alpha^l \psi(0,\bfz_1)A_{\rm cl}^l(\bfZ_1)\nonumber\\
&&\times \phi^{\dag}(0,\bfZ_1)\phi(0,\bfZ_1)
\phi^{\dag}(t,\bfX)\psi^{\dag}(t,\bfx)
\int d\bfx_1\, d\bfX_1 \psi^{\dag}(t_1,\bfx_1)\nonumber\\
&&\times \alpha^i \psi(t_1,\bfx_1)A^i(t_1,\bfx_1+\bfX_1)
\phi^{\dag}(t_1,\bfX_1)\phi(t_1,\bfX_1)\nonumber\\
&&\times \int d\bfY_1\; A^k(t_2,\bfY_1)
\phi^{\dag}(t_2,\bfY_1) (-i\nabla^k_{\bfY_1})
\phi(t_2,\bfY_1)|0\ra_{\rm con} \psi_a(\bfx)\,,
\ee
where we have used equation (\ref{timeord}).
To evaluate this expression we employ
the Wick theorem and equation (\ref{eq61})
for the nucleus propagator.
Keeping only the terms which will contribute
to the state-dependent energy shift,
 we obtain
\be \label{eq77}
\Delta g_a(E',E,\bfK',\bfK)&=&
-\frac{e^2|e|Z}{M}\int_{-\infty}^{\infty} dt\,dt'\,
dt_1\, dt_2\;
\int d\bfx\, d\bfx'\, d\bfx_1\, d\bfz_1\; \exp{(iE't'-iEt)}
\nonumber\\
&&\times [\psi_a^{\dag}(\bfx')S(t'-t_1,\bfx',\bfx_1)
\alpha^i S(t_1,\bfx_1,\bfz_1)\alpha^l S(-t,\bfz_1,\bfx)
\psi_a(\bfx)\nonumber\\
&&+\psi_a^{\dag}(\bfx')S(t',\bfx',\bfz_1)
\alpha^l S(-t_1,\bfz_1,\bfx_1)\alpha^i S(t_1-t,\bfx_1,\bfx)
\psi_a(\bfx)]\nonumber\\
&&\times D^{ik}(t_1-t_2,\bfx_1)
[\theta(t'-t_1)\theta(t_1-t_2)\theta(t_2)\theta(-t)
B_1^{kl}(\bfK',\bfK)\nonumber\\
&&+\theta(t'-t_1)\theta(t_1)\theta(-t_2)\theta(t_2-t)
B_2^{kl}(\bfK',\bfK)\nonumber\\
&&+\theta(t'-t_2)\theta(t_2-t_1)\theta(t_1)\theta(-t)
B_1^{kl}(\bfK',\bfK)\nonumber\\
&&+\theta(t'-t_2)\theta(t_2)\theta(-t_1)\theta(t_1-t)
B_1^{kl}(\bfK',\bfK)\nonumber\\
&&+\theta(t')\theta(-t_1)\theta(t_1-t_2)\theta(t_2-t)
B_2^{kl}(\bfK',\bfK)\nonumber\\
&&+\theta(t')\theta(-t_2)\theta(t_2-t_1)\theta(t_1-t)
B_2^{kl}(\bfK',\bfK)\,,
\ee
where
\be \label{eq78}
B_1^{kl}(\bfK',\bfK)&=&\frac{1}{(2\pi)^3}
\int d\bfX\;\exp{(-i\bfK'\cdot \bfX)}
(-i\nabla^k_{\bfX})A^l_{\rm cl}(\bfX)
\exp{(i\bfK\cdot \bfX)}\,, \\
B_2^{kl}(\bfK',\bfK)&=&\frac{1}{(2\pi)^3}
\int d\bfX\;\exp{(-i\bfK'\cdot \bfX)}
A^l_{\rm cl}(\bfX)(-i\nabla^k_{\bfX})
\exp{(i\bfK\cdot \bfX)}\,.
\label{eq79}
\ee
Integrating over the time variables, we obtain
\be \label{eq80}
\Delta g_a(E',E,\bfK',\bfK)&=&
-\frac{e^2|e|Z}{M}\frac{i}{2\pi}\int_{-\infty}^{\infty}
d \omega \; \sum_n \la a|\alpha^i D^{ik}(\omega)|n\ra
\la n|\alpha^l|a\ra \nonumber\\
&&\times \Bigl\{\theta(\veps_n)
\frac{i}{E'-\veps_a}\,\frac{i}{E'-\omega-\veps_n+i0}\,
\frac{i}{E'-\veps_n}\,\frac{i}{E-\veps_a}\,B_1^{kl}(\bfK',\bfK)
\nonumber\\
&&+\theta(\veps_n)\frac{i}{E'-\veps_a}\,\frac{i}{E'-\omega-\veps_n+i0}\,
\frac{i}{E-\omega-\veps_a+i0}\,\frac{i}{E-\veps_a}\,\nonumber\\
&&\times B_2^{kl}(\bfK',\bfK)
+\theta(\veps_n)\frac{i}{E'-\veps_a}\,\frac{i}{E'+\omega-\veps_a+i0}\,
\nonumber\\
&&\times \frac{i}{E'-\veps_n}\,\frac{i}{E-\veps_a}\,B_1^{kl}(\bfK',\bfK)
\nonumber\\
&&-\theta(-\veps_n)\frac{i}{E'-\veps_a}\,\frac{i}{E'+\omega-\veps_a+i0}\,
\frac{i}{E+\omega+\veps_n-2\veps_a+i0}\,\frac{i}{E-\veps_a}
\,\nonumber\\
&& \times B_1^{kl}(\bfK',\bfK)
-\theta(-\veps_n)\frac{i}{E'-\veps_a}\,\frac{i}{E-\omega-\veps_a+i0}\,
\nonumber\\
&&\times 
\frac{i}{E+\veps_n-2\veps_a}\,\frac{i}{E-\veps_a}\,B_2^{kl}(\bfK',\bfK)
\nonumber\\
&&-\theta(-\veps_n)
\frac{i}{E'-\veps_a}\,\frac{i}{E+\omega+\veps_n-2\veps_a+i0}\,
\frac{i}{E+\veps_n-2\veps_a}\,\frac{i}{E-\veps_a}\,
\nonumber\\
&&\times B_2^{kl}(\bfK',\bfK)
\Bigr\}\nonumber\\
&&-\frac{e^2|e|Z}{M}\frac{i}{2\pi}\int_{-\infty}^{\infty}
d \omega \; \sum_n \la a|\alpha^l|n\ra
\la n|\alpha^iD^{ik}(\omega)|a\ra \nonumber\\
&&\times \Bigl\{
-\theta(-\veps_n)
\frac{i}{E'-\veps_a}\,\frac{i}{E'-\omega+\veps_n-2\veps_a+i0}\,
\frac{i}{E'+\veps_n-2\veps_a}\,\frac{i}{E-\veps_a}\,
\nonumber\\
&&\times B_1^{kl}(\bfK',\bfK)
-\theta(-\veps_n)\frac{i}{E'-\veps_a}\,\frac{i}{E-\omega-\veps_a+i0}\,
\nonumber\\
&&\times
\frac{i}{E'-\omega+\veps_n-2\veps_a+i0}\,\frac{i}{E-\veps_a}
\,B_2^{kl}(\bfK',\bfK)
\nonumber\\
&&-\theta(-\veps_n)\frac{i}{E'-\veps_a}\,\frac{i}{E'+\omega-\veps_a+i0}\,
\frac{i}{E'+\veps_n-2\veps_a}\,\frac{i}{E-\veps_a}\,
\nonumber\\
&&\times B_1^{kl}(\bfK',\bfK)
+\theta(\veps_n)\frac{i}{E'-\veps_a}\,\frac{i}{E'+\omega-\veps_a+i0}\,
\nonumber\\
&&\times\frac{i}{E+\omega-\veps_n+i0}\,\frac{i}{E-\veps_a}
\,B_1^{kl}(\bfK',\bfK)
\nonumber\\
&&+\theta(\veps_n)\frac{i}{E'-\veps_a}\,\frac{i}{E-\omega-\veps_a+i0}\,
\frac{i}{E-\veps_n}\,\frac{i}{E-\veps_a}\,
\nonumber\\
&&\times
B_2^{kl}(\bfK',\bfK)
+\theta(\veps_n)\frac{i}{E'-\veps_a}\,\frac{i}{E+\omega-\veps_n+i0}\,
\nonumber\\
&&\times
\frac{i}{E-\veps_n}\,\frac{i}{E-\veps_a}\,B_2^{kl}(\bfK',\bfK)
\Bigr\}\,.
\ee
Taking into account that $\la n|\alpha^l|a\ra=0$ if
$\veps_n=\veps_a$, we have
\be \label{eq81}
\lefteqn{\frac{1}{2\pi i}  \oint_{\Gamma} dE\,
\frac{1}{2\pi i}
        \oint_{\Gamma'} dE'\;
\Delta g_{a}(E',E,\bfK',\bfK)\;\;\;\;\;\;\;\;\;\;\;\;\;\;\;\;\;
\;\;\;\;\;\;\;\;\;\;\;\;\;\;\;\;\;\;\;\;\;\;\;\;\;\;\;\;\;}
\nonumber\\
&=&-\frac{e^2|e|Z}{2M}\sum_n^{\veps_n\ne \veps_a}
\Bigl(\frac{\la a|\alpha^i D^{ik}(0)|n\ra \la n|
\alpha^l|a\ra}{\veps_a-\veps_n}\nonumber\\
&&+\frac{\la a|\alpha^l |n\ra \la n|
\alpha^i D^{ik}(0)|a\ra}{\veps_a-\veps_n}\Bigr)[B_1^{kl}(\bfK',\bfK)
+B_2^{kl}(\bfK',\bfK)]
\nonumber\\
&&-\frac{e^2|e|Z}{M}\frac{i}{2\pi}\int_{-\infty}^{\infty}
d\omega\;
\sum_n^{\veps_n\ne \veps_a}
\Bigl(\frac{\la a|\alpha^i D^{ik}(\omega)|n\ra \la n|
\alpha^l|a\ra}{(\veps_a-\veps_n)(\veps_a-\omega-\veps_n(1-i0))}
\nonumber\\
&&-\frac{\la a|\alpha^l |n\ra \la n|
\alpha^i D^{ik}(\omega)|a\ra}
{(\veps_a-\veps_n)
(\veps_a+\omega-\veps_n(1-i0))}\Bigr)C^{kl}(\bfK',\bfK)\,,
\ee
where
\be \label{eq82}
C^{kl}(\bfK',\bfK)=
B_1^{kl}(\bfK',\bfK)-B_2^{kl}(\bfK',\bfK)
=-\frac{i}{2}\,\eps_{kls}{\cal H}^s\delta(\bfK'-\bfK)\,
\ee
and $\eps_{kls}$ is the Levi-Civita symbol 
($\eps_{123}=\eps_{312}=\eps_{231}=1$,
$\eps_{321}=\eps_{132}=\eps_{213}=-1$, and
$\eps_{kls}=0$ if at least two of the indices are equal each other).
It can easily be shown that the term containing $[B_1^{kl}+B_2^{kl}]$
in equation (\ref{eq81}) is equal to zero.
Indeed, using the identity
\be \label{eq83}
\alpha^l=i[H_{\rm D},x^l]\,,
\ee
where $H_{\rm D}=-i\balpha \cdot\bnabla+\beta m +V$,
we obtain
\be \label{eq84}
\lefteqn{\sum_n^{\veps_n\ne \veps_a}
\Bigl(\frac{\la a|\alpha^i D^{ik}(0)|n\ra \la n|
\alpha^l|a\ra}{\veps_a-\veps_n}
+\frac{\la a|\alpha^l |n\ra \la n|
\alpha^i D^{ik}(0)|a\ra}{\veps_a-\veps_n}\Bigr)}
\nonumber\\
&=&i\sum_n^{\veps_n\ne \veps_a}
(-\la a|\alpha^i D^{ik}(0)|n\ra \la n|
x^l|a\ra+\la a|x^l |n\ra \la n|
\alpha^i D^{ik}(0)|a\ra)\nonumber\\
&=&i(-\la a|\alpha^i D^{ik}(0)x^l|a\ra+
\la a|x^l \alpha^i D^{ik}(0)|a\ra)=0\,.
\ee
Here we have taken into account that for the case
of an extended-charge nucleus considered here
there are no degenerate states of different parity.
Taking into account that to zeroth order the denominator
in equation (\ref{eq51}) is equal to 1, we have
\be \label{eq85}
\Delta E&=&\frac{e^2|e|Z}{M}\frac{i}{2}\,\eps_{kls}{\cal H}^s
\frac{i}{2\pi}\int_{-\infty}^{\infty}
d\omega\; \sum_n^{\veps_n\ne \veps_a}
\Bigl[\frac{\la a|\alpha^i D^{ik}(\omega)|n\ra \la n|
\alpha^l|a\ra}{(\veps_a-\veps_n)(\veps_a-\omega-\veps_n(1-i0))}
\nonumber\\
&&-\frac{\la a|\alpha^l |n\ra \la n|
\alpha^i D^{ik}(\omega)|a\ra}
{(\veps_a-\veps_n)(\veps_a+\omega-\veps_n(1-i0))}\Bigr]
\ee
Using identity (\ref{eq83}),
we obtain
\be \label{eq88}
\Delta E=\frac{e}{2M}\eps_{kls}{\cal H}^s
\frac{i}{2\pi}\int_{-\infty}^{\infty}
d\omega\; \{\la a|x^lG(\omega)D^k(\veps_a-\omega)|a\ra
+\la a|D^k(\veps_a-\omega)G(\omega)x^l|a\ra\}\,,
\ee
where $D^k(\omega)$ is defined by equation (\ref{eq68}).

The other contributions to $\Delta E^{(2b)}$ defined
by equation (\ref{eq54}) are derived in the same way.
As a result of this derivation,
keeping only the state-dependent contributions,
 we have
\be \label{eq89}
\Delta E=\frac{e}{2M}\eps_{kls}{\cal H}^s
\frac{i}{2\pi}\int_{-\infty}^{\infty}
d\omega\; \{\la a|x^lG(\omega)(m\alpha^k-p^k)|a\ra
+\la a|(m\alpha^k-p^k)G(\omega)x^l|a\ra\}\,.
\ee
The sum of expressions (\ref{eq88}) and (\ref{eq89})
gives the correction $\Delta E^{(2b)}$. 
 For the correction  $\Delta E^{(2)} \equiv
 \Delta E^{(2a)}+ \Delta E^{(2b)}$, where $\Delta E^{(2a)}$
is defined by equation (\ref{eq75}),
 we obtain
\be \label{eq90}
\Delta E^{(2)}&=&-\frac{e}{2M}\eps_{kls}{\cal H}^s
\frac{i}{2\pi}\int_{-\infty}^{\infty}
d\omega\; \{\la a|x^lG(\omega+\veps_a)
[p^k-D^k(\omega)]|a\ra
\nonumber\\
&&+\la a|[p^k-D^k(\omega)]G(\omega+\veps_a)x^l|a\ra\}\,.
\ee
For practical calculations, it is convenient
to represent this correction by the sum of the
lower-order and higher-order terms,
\be \label{eq91}
\Delta E^{(2)}&=&\Delta E_{\rm L}^{(2)}+\Delta E_{\rm H}^{(2)}\,,
\\
\Delta E_{\rm L}^{(2)}&=&
-\frac{e}{2M}\eps_{kls}{\cal H}^s
\{\la a|x^l[p^k-D^k(0)]|a\ra\,,
\label{eq92} \\
\Delta E_{\rm H}^{(2)}&=&-\frac{e}{2M}i\eps_{kls}{\cal H}^s
\frac{i}{2\pi}\int_{-\infty}^{\infty}
d\omega\;\frac{1}{\omega+i0}
 \{\la a|\alpha^lG(\omega+\veps_a)[p^k-D^k(\omega)]|a\ra
\nonumber\\
&&-\la a|[p^k-D^k(\omega)]G(\omega+\veps_a)\alpha^l|a\ra\}\,.
\label{eq93}
\ee
Here, as in equations (\ref{eq72})-(\ref{eq74}),
the term $\Delta E_{\rm L}^{(2)}$ contains the non-relativistic 
and lowest-order relativistic contributions and a part of
the higher-order relativistic contributions while the term
$\Delta E_{\rm H}^{(2)}$ contains all the higher-order
relativistic contributions
 which are not included into $\Delta E_{\rm L}^{(2)}$.

The total energy shift of first order in $m/M$ and of first order
in the interaction with the classical magnetic field is
$\Delta E^{({\rm tot})}=\Delta E^{(1)}+\Delta E^{(2)}$.
In the compact form, $\Delta E^{(1)}$ and $\Delta E^{(2)}$
are defined by equations (\ref{eq69}) and (\ref{eq90}),
respectively. For practical calculations,
it is more convenient to use the representation given
by equations (\ref{eq72})-(\ref{eq74}) and 
(\ref{eq91})-(\ref{eq93}).
To derive these equations, we have assumed that $V(\bfx)$
deviates from the pure Coulomb potential. However, since
the final formulas for the energy shift exhibit neither
singularities nor ambiguities when
$V(\bfx)\rightarrow V_{\rm C}=-\alpha Z/|\bfx|$,
the pure Coulomb potential can be used in these
formulas as well.

 The corresponding correction
to the bound-electron $g$ factor is defined as
\be \label{eq94}
\Delta g=\frac{\Delta E^{({\rm tot})}}{\mu_0{\cal H}m_j}\,,
\ee
where $\mu_0=|e|/(2m)$ is the Bohr magneton and
$m_j$ is the angular momentum projection of the state
under consideration. Here and below we assume that ${\bf {\cal H}}$
is directed along the $z$ axis.

\begin{center}
{\bf V.$\;$RECOIL CORRECTION TO THE BOUND-ELECTRON $g$ FACTOR 
TO LOWEST ORDERS IN $\alpha Z$}
\end{center}

To the lowest-order relativistic approximation,
the recoil correction to the $g$ factor is given by
\be \label{eq95}
\Delta E_{\rm L}^{({\rm tot})}=\Delta E_{\rm L}^{(1)}+
\Delta E_{\rm L}^{(2)}\,,
\ee
where  $\Delta E_{\rm L}^{(1)}$ and $\Delta E_{\rm L}^{(2)}$
are defined by equations (\ref{eq73}) and (\ref{eq92}),
respectively.  Let us calculate this correction
for an arbitrary state of a hydrogenlike atom.
For the case of the point-charge nucleus that we will consider,
this calculation can be performed analytically.

Consider first the calculation of $\Delta E_{\rm L}^{(1)}$.
According to equation (\ref{eq73}), it is
\be \label{eq96}
\Delta E_{\rm L}^{(1)}=\frac{1}{M} \la \delta a|
\Bigr[
\bfp^2-\frac{\alpha Z}{r}(\balpha\cdot\bfp
+(\balpha\cdot \bfn)(\bfn\cdot\bfp)\Bigr]|a\ra\,,
\ee
where $\bfn=\bfr/r$. Taking into account that 
$\bfp^2=(\balpha \cdot \bfp)^2$ and
$(\balpha\cdot \bfp)=H_{\rm D}-\beta m-V_{\rm C}$,
one easily obtains
\be \label{eq97}
\la \delta a|\bfp^2|a\ra &=&\la \delta a|
(\veps_a+\beta m-V_{\rm C})(\veps_a-\beta m-V_{\rm C})|a\ra
\nonumber\\
&&+i \la \delta a|(\balpha \cdot\bnabla V_{\rm C})|a\ra\,.
\ee
The second term in equation (\ref{eq96}) can be transformed
as (see, e.g., \cite{sha94})
\be \label{eq98}
-\la \delta a|\frac{\alpha Z}{r}[\balpha\cdot \bfp
+(\balpha \cdot \bfn)(\bfn \cdot \bfp)]|a\ra 
&=&-\la \delta a|\frac{\alpha Z}{r}\Bigr[2\veps_a-2\beta m
-2V_{\rm C}\nonumber\\
&&+\frac{i}{r}(\balpha \cdot \bfn)(\beta \kappa+1)
\Bigr]|a\ra\,,
\ee
where $\kappa=(-1)^{j+l+1/2}(j+1/2)$ is the relativistic
angular quantum number of the state $a$,
$j$ is the total angular momentum, and $l=j\pm 1/2$
defines the parity of the state.
The wave function correction $|\delta a\ra$ defined by equation
(\ref{eq71}) can easily be found analytically using
the method of the generalized  virial relations for
the Dirac equation developed in \cite{sha91}.
Since the operator sandwiched between $|a\ra$ and $|\delta a\ra$
in equation for $\Delta E_{\rm L}^{(1)}$ conserves
the angular quantum numbers, we need only that component of
 $|\delta a\ra$ which has the same angular quantum numbers
as the unperturbed state  $|a\ra$.
This component is
\be \label{eq99}
|\delta a\ra_{\kappa m_j}=
\left(\begin{array}{c}
X(r)\Omega_{\kappa m_j}({\bf n})\\
iY(r)\Omega_{-\kappa m_j}({\bf n})
\end{array}\right)\;,
\ee
where
\be \label{eq100}
X(r)&=&b_0\Bigl[\frac{2m\kappa -m+2\kappa \veps_a}{2m^2}r
+\frac{\alpha Z}{m^2}\kappa\Bigr]f(r)
+\frac{\kappa-2\kappa^2}{2m^2}g(r)\,,\\
Y(r)&=&b_0\Bigl[\frac{2m\kappa +m-2\kappa \veps_a}{2m^2}r
-\frac{\alpha Z}{m^2}\kappa\Bigr]g(r)
+\frac{\kappa+2\kappa^2}{2m^2}f(r)\,,
\label{eq101}\\
b_0&=&-\frac{e}{2}{\cal H}\frac{\kappa}{j(j+1)}m_j\,,
\label{eq102}
\ee
$g(r)$ and $f(r)$ are the radial parts of the unperturbed
wave function defined as
\be \label{eq103}
|a\ra=
\left(\begin{array}{c}
g(r)\Omega_{\kappa m_j}({\bf n})\\
if(r)\Omega_{-\kappa m_j}({\bf n})
\end{array}\right)\;.
\ee
Integrating over the angular variables in 
equations (\ref{eq97}) and (\ref{eq98}), we find
\be \label{eq104}
\Delta E_{\rm L}^{(1)}&=&\frac{b_0}{M}\int_{0}^{\infty}dr\;
r^2\Bigl\{X(r)g(r)[-2V_{\rm C}m-V_{\rm C}^2+\veps_a^2-m^2]
\nonumber\\
&&+Y(r)f(r)[2V_{\rm C}m-V_{\rm C}^2+\veps_a^2-m^2]\nonumber\\
&&+[X(r)f(r)+Y(r)g(r)]\frac{\alpha Z}{r^2}\kappa\Bigr\}\,.
\ee
Substituting expressions (\ref{eq100}) and (\ref{eq101})
into equation (\ref{eq104}), we obtain
\be \label{eq105}
\Delta E_{\rm L}^{(1)}&=&\frac{b_0}{M}\Bigl\{
\alpha Z\frac{2\kappa \veps_a-m}{m}C^0+
(\alpha Z)^2 \frac{\kappa}{m}C^{-1}+
(\veps_a^2-m^2)\frac{\kappa}{m}C^{1}\nonumber\\
&&+\alpha Z \frac{\kappa^2}{2m^2}C^{-2}+(\veps_a^2-m^2)
\frac{\kappa}{2m^2}A^0-\alpha Z\frac{\kappa^2}{m}A^{-1}
\nonumber\\
&&-(\alpha Z)^2\frac{\kappa}{2m^2}A^{-2}
-(\veps_a^2-m^2)\frac{\kappa^2}{m^2}B^0
+\alpha Z\frac{3m\kappa-2\kappa^2 \veps_a}{2m^2}B^{-1}\Bigr\}\,,
\ee
where we have used the notations \cite{sha91}
\be \label{eq106}
A^s=\int_{0}^{\infty}dr\; r^{2+s}(g^2+f^2)\,,\;\;\;\;\;\;
B^s=\int_{0}^{\infty}dr\; r^{2+s}(g^2-f^2)\,,\;\;\;\;\;\;
C^s=2\int_{0}^{\infty}dr\; r^{2+s}gf\,.
\ee
The integrals $A^s$, $B^s$, and $C^s$ are easily calculated 
by the recurrent equations given in \cite{sha91}
(the relevant equations were first derived in \cite{eps62}).
We have
\be \label{eq107}
C^0&=&\frac{\kappa}{\alpha Z}\frac{m^2-\veps_a^2}{m^2}\,,\;\;\;\;\;\;\;\;
C^{-1}=\frac{(\alpha Z)^2\kappa m}{N^3\gamma}\,,\;\;\;\;\;\;\;\;
C^1=\frac{2\kappa\veps_a-m}{2m^2}\,,\;\;\;\;\;\;\;\;
\nonumber\\
C^{-2}&=&\frac{2(\alpha Z)^3[2\kappa(\gamma+n_r)-N]}{
N^4(4\gamma^2-1)\gamma}\,,\;\;\;\;\;\;\;\;
A^0=1\,,\;\;\;\;\;\;\;\;
A^{-1}=\frac{\alpha Z m(\kappa^2+n_r\gamma)}{N^3 \gamma}\,,\;\;\;\;\;\;\;\;
\nonumber\\
A^{-2}&=&\frac{2(\alpha Z)^2 \kappa[2\kappa(\gamma+n_r)-N]m^2}
{N^4(4\gamma^2-1)\gamma}\,,\;\;\;\;\;\;\;\;
B^0=\frac{\veps_a}{m}\,,\;\;\;\;\;\;\;\;
B^{-1}=\frac{m^2-\veps_a^2}{\alpha Z m}\,,
\ee
where $n_r$ is the radial quantum number,
$N=\sqrt{n^2-2n_r(|\kappa|-\gamma)}$, and
$n=n_r+|\kappa|$ is the principal quantum number.
Substituting formulas (\ref{eq107}) into equation
(\ref{eq105}), we obtain
\be \label{eq108}
\Delta E_{\rm L}^{(1)}=-\frac{e}{2}\,\frac{{\cal H}}{M}\,
\frac{\kappa^2}{2j(j+1)}\frac{m^2-\veps_a^2}{m^2}m_j\,.
\ee

Consider now the term $\Delta E_{\rm L}^{(2)}$.
According to equation (\ref{eq92}), we have
\be \label{eq109}
\Delta E_{\rm L}^{(2)}&=&\frac{e}{2}\,\frac{{\cal H}}{M}\,
\la a|([\bfr\times \bfp]_z-[\bfr\times \bfD(0)]_z)|a\ra
\nonumber\\
&=&\frac{e}{2}\,\frac{{\cal H}}{M}\,
\la a|\Bigl(l_z-\frac{\alpha Z}{2r}[\bfr\times \balpha]_z\Bigr)|a\ra\,.
\ee
Integrating over the angular and radial variables, we obtain
\be \label{eq110}
\Delta E_{\rm L}^{(2)}&=&\frac{e}{2}\,\frac{{\cal H}}{M}\,
\frac{1}{2j(j+1)}
\Bigl\{j(j+1)-\frac{3}{4}+l(l+1)\frac{m+\veps_a}{2m}
\nonumber\\
&&+(2j-l)(2j-l+1)\frac{m-\veps_a}{2m}
-\kappa^2\frac{m^2-\veps_a^2}{m^2}\Bigr\}m_j\,.
\ee
For the sum $\Delta E_{\rm L}=\Delta E_{\rm L}^{(1)}+
\Delta E_{\rm L}^{(2)}$, we find
\be \label{eq111}
\Delta E_{\rm L}=\frac{e}{2}\,\frac{{\cal H}}{M}\,
\frac{2\kappa^2\veps_a^2+\kappa m \veps_a-m^2}{2m^2 j(j+1)}m_j\,.
\ee
The related contribution to the $g$ factor is
\be \label{eq112}
\Delta g_{\rm L}=-\frac{m}{M}\,
\frac{2\kappa^2\veps_a^2+\kappa m \veps_a-m^2}{2m^2 j(j+1)}\,.
\ee
To the two lowest orders in $\alpha Z$, we have
\be \label{eq113}
\Delta g_{\rm L}=-\frac{m}{M}\,
\frac{1}{j(j+1)}\Bigl[\kappa^2+\frac{\kappa}{2}
-\frac{1}{2}-\Bigl(\kappa^2+\frac{\kappa}{4}\Bigr)
\frac{(\alpha Z)^2}{n^2}\Bigr]\,.
\ee
For an $ns$ state, formula (\ref{eq112}) yields
\be \label{eq114}
\Delta g_{\rm L}=-\frac{m}{M}\,\frac{2}{3}\,
\frac{2\veps_a^2- m \veps_a-m^2}{m^2}\,,
\ee
while formula (\ref{eq113}) gives
\be \label{eq115}
\Delta g_{\rm L}=\frac{m}{M}\frac{(\alpha Z)^2}{n^2}\,.
\ee
Formula (\ref{eq115}) agrees with the related result obtained
for the $1s$ state in \cite{fau70a,fau70b,gro70,gro71}.

$$
\;
$$
$$
\;
$$

\begin{center}
{\bf VI.$\;$CONCLUSION}
\end{center}

In this paper we formulated the systematic QED method
for calculations of the nuclear recoil corrections to
the energy levels and to the electronic $g$ factor 
in atoms to first order in $m/M$ and to all orders in $\alpha Z$.
Employing this method, we derived the complete 
$\alpha Z$-dependence expression for the recoil correction
to the bound-electron $g$ factor in a hydrogenlike
atom (equations (\ref{eq69}),{\ref{eq90})).
 This expression was also represented as the sum
of the lower-order and higher-order terms (equations
(\ref{eq72})-(\ref{eq74}), (\ref{eq91})-(\ref{eq93})).
The lower-order
term contains all the non-relativistic and lowest-order
relativistic contributions and a part of the higher-order
corrections. For an arbitrary state of a hydrogenlike atom,
a simple analytical formula was derived for this term
(equation (\ref{eq111})). As to the higher-order
term,  we expect that its numerical evaluation 
can be performed in the same way
as the evaluation of
 the related contribution to the energy levels
\cite{art95a,sha98c,sha98d}.
 This calculation,
which is equally important for low- and high-$Z$
systems, is under way and will be published elsewhere.

\begin{center}
{\bf ACKNOWLEDGEMENTS}
\end{center}

Valuable conversations with D. Arbatsky, T. Beier,
S. Karshenboim, J. Kluge, W. Quint, and V. Yerokhin 
are gratefully acknowledged. I am grateful to the
Atomic group of GSI and personally to T. Beier
and J. Kluge for hospitality during the visit in winter 2001. 
This work was supported in part by RFBR (Grant No. 
01-02-17248) and by the program
"Russian Universities - Basic Research" (project No. 3930).

\end{document}